\documentclass[twocolumn,showpacs,preprintnumbers,amsmath,amssymb,eqsecnum]{revtex4}
\usepackage{dcolumn}
\usepackage{bm}
\usepackage[dvips]{graphicx}
\usepackage{wrapfig}
\usepackage{psfrag}
\begin{document}

\def\sh{\mathop{\rm sh}\nolimits}
\def\ch{\mathop{\rm ch}\nolimits}
\def\var{\mathop{\rm var}}\def\exp{\mathop{\rm exp}\nolimits}
\def\Re{\mathop{\rm Re}\nolimits}
\def\Sp{\mathop{\rm Sp}\nolimits}
\def\kp{\mathop{\text{\ae}}\nolimits}
\def\bk{{\bf {k}}}
\def\bp{{\bf {p}}}
\def\bq{{\bf {q}}}
\def\lra{\mathop{\longrightarrow}}
\def\Const{\mathop{\rm Const}\nolimits}
\def\sh{\mathop{\rm sh}\nolimits}
\def\ch{\mathop{\rm ch}\nolimits}
\def\var{\mathop{\rm var}}
\def\mK{\mathop{{\mathfrak {K}}}\nolimits}
\def\mR{\mathop{{\mathfrak {R}}}\nolimits}
\def\mv{\mathop{{\mathfrak {v}}}\nolimits}
\def\mV{\mathop{{\mathfrak {V}}}\nolimits}
\def\mD{\mathop{{\mathfrak {D}}}\nolimits}
\def\mN{\mathop{{\mathfrak {N}}}\nolimits}
\def\mS{\mathop{{\mathfrak {S}}}\nolimits}

\newcommand\ve[1]{{\mathbf{#1}}}

\def\Re{\mbox {Re}}
\newcommand{\Z}{\mathbb{Z}}
\newcommand{\R}{\mathbb{R}}
\def\mC{\mathop{{\mathfrak {C}}}\nolimits}
\def\mK{\mathop{{\mathfrak {K}}}\nolimits}
\def\mk{\mathop{{\mathfrak {k}}}\nolimits}
\def\mR{\mathop{{\mathfrak {R}}}\nolimits}
\def\mv{\mathop{{\mathfrak {v}}}\nolimits}
\def\mV{\mathop{{\mathfrak {V}}}\nolimits}
\def\mD{\mathop{{\mathfrak {D}}}\nolimits}
\def\mN{\mathop{{\mathfrak {N}}}\nolimits}
\def\ml{\mathop{{\mathfrak {l}}}\nolimits}
\def\mf{\mathop{{\mathfrak {f}}}\nolimits}
\def\mg{\mathop{{\mathfrak {g}}}\nolimits}
\newcommand{\ccm}{{\cal M}}
\newcommand{\cE}{{\cal E}}
\newcommand{\cV}{{\cal V}}
\newcommand{\cI}{{\cal I}}
\newcommand{\cR}{{\cal R}}
\newcommand{\cK}{{\cal K}}
\newcommand{\cH}{{\cal H}}
\newcommand{\cW}{{\cal W}}

\def\br{\mathop{{\bf {r}}}\nolimits}
\def\bS{\mathop{{\bf {S}}}\nolimits}
\def\bA{\mathop{{\bf {A}}}\nolimits}
\def\bJ{\mathop{{\bf {J}}}\nolimits}
\def\bn{\mathop{{\bf {n}}}\nolimits}
\def\bg{\mathop{{\bf {g}}}\nolimits}
\def\bv{\mathop{{\bf {v}}}\nolimits}
\def\be{\mathop{{\bf {e}}}\nolimits}
\def\bp{\mathop{{\bf {p}}}\nolimits}
\def\bz{\mathop{{\bf {z}}}\nolimits}
\def\bbf{\mathop{{\bf {f}}}\nolimits}
\def\bb{\mathop{{\bf {b}}}\nolimits}
\def\ba{\mathop{{\bf {a}}}\nolimits}
\def\bx{\mathop{{\bf {x}}}\nolimits}
\def\by{\mathop{{\bf {y}}}\nolimits}
\def\br{\mathop{{\bf {r}}}\nolimits}
\def\bs{\mathop{{\bf {s}}}\nolimits}
\def\bH{\mathop{{\bf {H}}}\nolimits}
\def\bk{\mathop{{\bf {k}}}\nolimits}
\def\be{\mathop{{\bf {e}}}\nolimits}
\def\bnul{\mathop{{\bf {0}}}\nolimits}
\def\bq{{\bf {q}}}

\newcommand{\oV}{\overline{V}}
\newcommand{\vkp}{\varkappa}
\newcommand{\os}{\overline{s}}
\newcommand{\opsi}{\overline{\psi}}
\newcommand{\ov}{\overline{v}}
\newcommand{\oW}{\overline{W}}
\newcommand{\oPhi}{\overline{\Phi}}

\def\mI{\mathop{{\mathfrak {I}}}\nolimits}
\def\mA{\mathop{{\mathfrak {A}}}\nolimits}

\def\st{\mathop{\rm st}\nolimits}
\def\tr{\mathop{\rm tr}\nolimits}
\def\sign{\mathop{\rm sign}\nolimits}
\def\d{\mathop{\rm d}\nolimits}
\def\const{\mathop{\rm const}\nolimits}
\def\O{\mathop{\rm O}\nolimits}
\def\Spin{\mathop{\rm Spin}\nolimits}
\def\exp{\mathop{\rm exp}\nolimits}
\def\SU{\mathop{\rm SU}\nolimits}
\def\mU{\mathop{{\mathfrak {U}}}\nolimits}
\newcommand{\cU}{{\cal U}}
\newcommand{\cD}{{\cal D}}

\def\mI{\mathop{{\mathfrak {I}}}\nolimits}
\def\mA{\mathop{{\mathfrak {A}}}\nolimits}
\def\mU{\mathop{{\mathfrak {U}}}\nolimits}

\def\st{\mathop{\rm st}\nolimits}
\def\tr{\mathop{\rm tr}\nolimits}
\def\sign{\mathop{\rm sign}\nolimits}
\def\d{\mathop{\rm d}\nolimits}
\def\const{\mathop{\rm const}\nolimits}
\def\O{\mathop{\rm O}\nolimits}
\def\Spin{\mathop{\rm Spin}\nolimits}
\def\exp{\mathop{\rm exp}\nolimits}

\title{A Note on the Possible Existence of an Instanton-like Self-Dual Solution to Lattice Euclidean Gravity}

\author {S.N. Vergeles\vspace*{4mm}\footnote{{e-mail:vergeles@itp.ac.ru}}}

\affiliation{Landau Institute for Theoretical Physics,
Russian Academy of Sciences,
Chernogolovka, Moscow region, 142432 Russia \linebreak
and   \linebreak
Moscow Institute of Physics and Technology, Department
of Theoretical Physics, Dolgoprudnyj, Moskow region,
141707 Russia}

\begin{abstract} The self-dual solution to lattice Euclidean gravity is constructed.
In contrast to the well known Eguchi-Hanson solution to continuous Euclidean
Gravity, the lattice solution is asymptotically {\it{globally}} Euclidean, i.e., the 
boundary of the space as $r\longrightarrow\infty$ is $S^3=SU(2)$. 
\end{abstract}

\pacs{11.15.-q, 11.15.Ha}

\maketitle

\section{Introduction}

Soon after the discovery of the self-dual (instanton) solution to the 4D Euclidean Yang-Mills theory
\cite{1}, the self-dual solution to 4D Euclidean Gravity has been obtained \cite{2}, \cite{3} (see
also \cite{4}, \cite{5}, \cite{6}). But the space-time boundary as $r\longrightarrow\infty$ is $S^3/{\Z}_2$ and not $S^3$
in the case of Eguchi-Hanson gravitational instanton, in contrast to the Yang-Mills one.
Otherwise, a "cone-tipe" singularity (effective delta-function in the
curvature at the instanton centre for $r=a$) would be necessary.
This means that the space-time topology of the gravitational instanton solution
differs crucially from the topology of a real space-time. Therefore, though the action of the Eguchi-Hanson solution is zeroth, the physical meaning of the solution is not clear.

I construct here the analogue of the Eguchi-Hanson self-dual solution to the 4D Euclidean lattice Gravity
with zeroth action. The solution transforms locally into the Eguchi-Hanson solution as $r\longrightarrow\infty$.
The reason is that the considered lattice theory transforms into Einstein theory for a long-wavelength limit.
The remarkable fact is the discrete gravity self-dual solution wipes out a "cone-tipe" singularity 
at the center of instanton for the case the space-time boundary as $r\longrightarrow\infty$ is $S^3$.  Thus, then the gravitational instantons would exist if the real space-time exhibits the granularity property at super small scales. 

This article does not contain any new result in comparison with \cite{11}, but the main result is obtained
more rationally and transparently. The changes concerned only Section V.

A preliminary version of the work has been published in \cite{7}.

\section{Eguchi-Hanson self-dual solution}

First of all it is necessary to describe shortly the Eguchi-Hanson self-dual solution to continuous Euclidean
Gravity. Let $\gamma^a,\,\, \gamma^a\gamma^b+\gamma^b\gamma^a=2\delta^{ab}, \,\,a=1,2,3,4$ be the Hermitian Dirac matrices $4\times4$ in spinor representation
($\sigma^{\alpha},\,\,\alpha=1,2,3$ are Pauli matrices):
\begin{eqnarray}
\gamma^{\alpha}=
 \left(
\begin{array}{cc}
0 & i\sigma^{\alpha}  \\
-i\sigma^{\alpha} & 0 \\
\end{array} \right),
\quad \gamma^4=
 \left(
\begin{array}{cc}
0 & 1  \\
1 & 0 \\
\end{array} \right), 
\nonumber \\
\gamma^5\equiv\gamma^1\gamma^2\gamma^3\gamma^4=
 \left(
\begin{array}{cc}
-1 & 0  \\
0 & 1 \\
\end{array} \right),  
\quad
\sigma^{ab}=\frac{1}{4}[\gamma^a,\gamma^b],
\nonumber \\
\sigma^{\alpha 4}= \frac{i}{2}\left(
\begin{array}{cc}
\sigma^{\alpha} & 0  \\
0 & -\sigma^{\alpha}  \\
\end{array} \right),  
\quad
\sigma^{\alpha\beta}= \frac{i\varepsilon_{\alpha\beta\gamma}}{2}\left(
\begin{array}{cc}
\sigma^{\gamma} & 0  \\
0 & \sigma^{\gamma}  \\
\end{array} \right).
\nonumber \\
{}
\label{inst10}
\end{eqnarray}
Let's consider the pure 4D Euclidean Gravity action in the Palatini form (the independent
variables are tetrad and connection): 
\begin{gather}
\mA=-\frac{1}{l^2_P}\int\tr\gamma^5{\it R}\wedge e\wedge e=
\nonumber \\
=-\frac{1}{l^2_P}\int\left\{{\it R}^{\alpha}_{(+)}\wedge E^{\alpha}_{(+)}-
{\it R}^{\alpha}_{(-)}\wedge E^{\alpha}_{(-)}\right\},
\nonumber \\ 
 {\it R}\equiv 2(\d\omega+\omega\wedge\omega)
= \frac{i\sigma^{\alpha}}{2} \left(
\begin{array}{cc}
{\it R}^{\alpha}_{(+)} & 0  \\
0 & {\it R}^{\alpha}_{(-)} \\
\end{array} \right),  
\nonumber \\
\omega\equiv\frac12\sigma^{ab}\omega^{ab}_{\mu}\d x^{\mu}=
 \frac{i\sigma^{\alpha}}{2} \left(
\begin{array}{cc}
\omega^{\alpha}_{(+)\mu} & 0  \\
0 & \omega^{\alpha}_{(-)\mu} \\
\end{array} \right)\d x^{\mu}, 
\nonumber \\
\omega^{\alpha}_{(\pm)}\equiv \left\{\pm\omega^{\alpha4}_{\mu}+\frac12\varepsilon_{\alpha\beta\gamma}
\omega^{\beta\gamma}_{\mu}\right\}\d x^{\mu},
\nonumber \\
{\it R}^{\alpha}_{(\pm)}=2\d\omega^{\alpha}_{(\pm)}-
\varepsilon_{\alpha\beta\gamma}\omega^{\beta}_{(\pm)}\wedge\omega^{\gamma}_{(\pm)},  
\nonumber \\
e\equiv\gamma^ae^a_{\mu}\d x^{\mu},
\nonumber \\
E^{\alpha}_{(\pm)}\equiv\left\{\pm\left(e^{\alpha}_{\lambda}e^4_{\rho}-e^4_{\lambda}e^{\alpha}_{\rho}\right)+
\varepsilon_{\alpha\beta\gamma}e^{\beta}_{\lambda}e^{\gamma}_{\rho}\right\}\d x^{\lambda}\wedge\d x^{\rho}. 
\label{inst20}
\end{gather}
One can take six 1-forms $\omega^{\alpha}_{(\pm)}$ as independent variables instead of
six 1-forms $\omega^{ab}$. Obviously, the representation (\ref{inst20}) is consistent with
the representation of the group $\Spin(4)\approx\Spin(4)_{(+)}\otimes\Spin(4)_{(-)}\approx
\SU(2)_{(+)}\otimes\SU(2)_{(-)}$.

The following equations are equivalent
\begin{gather}
\omega=\pm\gamma^5\omega \longleftrightarrow \omega^{ab}=\mp\frac12\varepsilon_{abcd}\omega^{cd}
\longleftrightarrow \omega^{\alpha}_{(\pm)}=0.
\label{inst40}
\end{gather}
Eqs. (\ref{inst40}) imply the following one:
\begin{gather}
{\it R}=\pm\gamma^5{\it R} \longleftrightarrow {\it R}^{ab}=\mp\frac12\varepsilon_{abcd}{\it R}^{cd}
\longleftrightarrow {\it R}^{\alpha}_{(\pm)}=0.
\label{inst50}
\end{gather}
The action stationarity condition relative to the connection gives the equation
\begin{gather}
\delta\mA/\delta\omega^{ab}_{\mu}=0 
\longrightarrow  \d e^a+\omega^{ab}\wedge e^b=0,
\label{inst60}
\end{gather}
which determines uniquely the connection forms for fixed forms $e^a$.
Additionally, Eq. (\ref{inst60}) imply the algebraic Bianchi identity for Riemannian tensor,
the combination of which with Eq. (\ref{inst50}) leads to the Einstein's equation
\begin{gather}
{\it R}_{ab}\equiv{\it R}^c_{acb}=0.
\label{inst80}
\end{gather}
On the other hand, Einstein's equation is equivalent to the action stationarity condition relative to the
forms $e^a_{\mu}$:
\begin{gather}
{\it R}_{ab}=0\longleftrightarrow  \delta\mA/\delta e^a_{\mu}=0.
\label{inst90}
\end{gather}
The question arises: why the additional Eq. (\ref{inst40}) does not come into conflict with Eqs. (\ref{inst60}) and  (\ref{inst90})?  To answer this question let's consider the case with the lower sign in Eq. (\ref{inst40}) when
\begin{gather}
\omega^{\alpha}_{(-)\mu}=0. 
\label{inst95}
\end{gather}
The combination of the part of Eqs. (\ref{inst60}) $\delta\mA/\delta\omega^{\alpha}_{(-)\mu}=0$ 
and Eq. (\ref{inst95}) gives the following 12 equations:
\begin{gather}
\varepsilon^{\mu\nu\lambda\rho}\partial_{\nu}E^{\alpha}_{(-)\lambda\rho}=0.
\label{inst100}
\end{gather}
Now we must solve the system of equations (\ref{inst90}), (\ref{inst95}), (\ref{inst100}) and
\begin{gather}
\delta\mA/\delta\omega^{\alpha}_{(+)\mu}=0.
\label{inst104}
\end{gather}

Note that Eqs. (\ref{inst60}) and (\ref{inst90}) do not fix the variables
$\omega^{ab}_{\mu}, \ e^a_{\mu}$ completely but up to the gauge (orthogonal) transformations.
Here the gauge group leaves 6 unfixed functions.

Eqs. (\ref{inst100}) do not fix the quantities $E^{\alpha}_{(-)\lambda\rho}$ completely but
up to summands of the forme $\left(\partial_{\lambda}\Psi^{\alpha}_{(-)\rho}-\partial_{\rho}\Psi^{\alpha}_{(-)\lambda}\right)$
where $\Psi^{\alpha}_{(-)\lambda}$ are 3 arbitrary vector fields (12 functions altogether).
But each of three vector fields $\Psi^{\alpha}_{(-)\lambda}$ contains only 3 independent functions 
due to invariance of the expression  $\left(\partial_{\lambda}\Psi^{\alpha}_{(-)\rho}-\partial_{\rho}\Psi^{\alpha}_{(-)\lambda}\right)$
relative to the changes $\Psi^{\alpha}_{(-)\lambda}\longrightarrow \Psi^{\alpha}_{(-)\lambda}+
\partial_{\lambda}\phi^{\alpha}_{(-)}$. As a result, Eqs. (\ref{inst100}) fix no more than additional $12-3\times3=3$
functions. This means that the gauge subgroup $\Spin(4)_{(-)}$ is broken by Eq. (\ref{inst95}).
So we see that the system of equations (\ref{inst90})-(\ref{inst104}) is consistent,
though it fixes the gauge subgroup $\Spin(4)_{(-)}$. The Eguchi-Hanson solution is the simplest solution of the system.
Let's write out this solution \cite{2}, \cite{3}.

Let $x^i=(r,\,\theta,\,\varphi,\,\psi)$, where $(\theta,\,\varphi,\,\psi)$ be the Euler angles.
The cartesian coordinates $x^{\mu}$ in ${\boldsymbol{\R^4}}$ are connected with the coordinates $x^i$
as follows:
\begin{gather}
z_1\equiv x^1+ix^2=r\cos\frac{\theta}{2}\exp\left[\frac{i}{2}(\psi+\varphi)\right],
\nonumber \\
z_2\equiv x^3+ix^4=r\sin\frac{\theta}{2}\exp\left[\frac{i}{2}(\psi-\varphi)\right].
\label{inst108}
\end{gather}
There is a one-to-one correspondence between these two coordinate systems if the Euler angles vary in the ranges
\begin{gather}
0\leq\theta\leq\pi, \quad 0\leq\varphi\leq 2\pi, \quad 0\leq\psi\leq 4\pi.
\label{inst130}
\end{gather}

The Eguchi-Hanson solution for the metrics $\d s^2\equiv \left(e^a_{\mu}\d x^{\mu}\right)^2$ has
the form
\begin{gather}
e^a_{\mu}\d x^{\mu}=
 \left(
\begin{array}{c}
\frac{r}{2}\left(\sin\psi\d\theta-\sin\theta\cos\psi\d\varphi\right)  \\
\frac{r}{2}\left(\cos\psi\d\theta+\sin\theta\sin\psi\d\varphi\right)  \\
-\frac{r}{2}g(r)\left(\cos\theta\d\varphi+\d\psi\right) \\
g(r)^{-1}\d r \\
\end{array} \right),
\nonumber \\
g(r)=\sqrt{1-\frac{a^4}{r^4}}, \quad r\geq a.
\label{inst110}
\end{gather}
For $a=0$, the metrics (\ref{inst110}) transforms to the 4D Euclidean metrics in Euler angles on
the $S^3$ with ranges (\ref{inst130}).
But  $0\leq\psi\leq 2\pi$ for the Eguchi-Hanson solution (when $a\neq0$) since the points with
coordinates $\psi$ and $(\psi+2\pi)$ and the same $(r,\,\theta,\,\varphi)$ are identified.
Otherwise, in order for the 
Chern-Gauss-Bonnet theorem to be satisfied in the case (\ref{inst130}), a "cone-tipe" singularity (effective delta-function in the curvature at the instanton centre for $r=a$) would be necessary (see \cite{3}).

The connection 1-forms are given by Eqs. (\ref{inst95}) and:
\begin{gather}
 \omega^1_{(+)}=\frac{2g}{r}e^1=g\left(\sin\psi\d\theta-\sin\theta\cos\psi\d\varphi\right), 
\nonumber \\
 \omega^2_{(+)}=\frac{2g}{r}e^2=g\left(\cos\psi\d\theta+\sin\theta\sin\psi\d\varphi\right),
\nonumber \\
\omega^3_{(+)}=2\left(\frac{2}{rg}-\frac{g}{r}\right)e^3=-\left(1+\frac{a^4}{r^4}\right)
\left(\cos\theta\d\varphi+\d\psi\right).
\label{inst140}
\end{gather}
We have 
\begin{gather}
\int_{r=\Const\rightarrow\infty}\left(\frac12\omega^1_{(+)}\right)\wedge\left(\frac12\omega^2_{(+)}\right)\wedge
\left(\frac12\omega^3_{(+)}\right)=-\pi^2
\label{inst143} 
\end{gather}
and
\begin{gather}
\int_{r\longrightarrow a+0}\left(\frac12\omega^1_{(+)}\right)\wedge\left(\frac12\omega^2_{(+)}\right)\wedge
\left(\frac12\omega^3_{(+)}\right)=0
\label{inst147}
\end{gather}
for the range $0\leq\psi\leq 2\pi$ and orientation $\theta,\,\varphi,\,\psi,\,r$.
The integral (\ref{inst143}) would be equal  to $(-\pi^2)$ for any $0<r=\Const<\infty$ in the case
$a=0$ (the Euclidean metrics in Euler angles). Thus, the boundary conditions (\ref{inst143})-(\ref{inst147})
determine the instanton Eguchi-Hanson solution with the same integration constant $a$ as in the relation
(\ref{inst147}). This means that the system of equations (\ref{inst90})-(\ref{inst104}) together with
the boundary conditions (\ref{inst143})-(\ref{inst147}) possess unique solution (\ref{inst110})
for the centrally symmetrical metrics anzats with the same integration constant $a$ as in the relation
(\ref{inst147}).

Note that 
\begin{gather}
\frac{i\sigma^{\alpha}}{2}\omega^{\alpha}_{(+)}=U^{-1}\d U \quad \mbox{as} \quad r\longrightarrow\infty,
\label{topol10}
\end{gather}
where
\begin{gather}
U=\exp\left(-\frac{i\sigma^3}{2}\varphi\right)\exp\left(\frac{i\sigma^2}{2}\theta\right)
\exp\left(-\frac{i\sigma^3}{2}\psi\right).
\label{topol20}
\end{gather}
According to the Eq. (\ref{topol10})
\begin{gather}
\frac{1}{12}\tr\left(U^{-1}\d U\right)\wedge\left(U^{-1}\d U\right)\wedge\left(U^{-1}\d U\right)=
\nonumber \\
=\left(\frac12\omega^1_{(+)}\right)\wedge\left(\frac12\omega^2_{(+)}\right)\wedge
\left(\frac12\omega^3_{(+)}\right) \quad \mbox{as} \quad r\longrightarrow\infty.
\label{topol30}
\end{gather}
Taking into account Eqs. (\ref{inst143}) and (\ref{topol30}), we obtain  for the angle ranges (\ref{inst130}):
\begin{gather}
\frac{1}{12}\tr\int\left(U^{-1}\d U\right)\wedge\left(U^{-1}\d U\right)\wedge\left(U^{-1}\d U\right)=-2\pi^2.
\label{topol40}
\end{gather}
This equality means that Eq. (\ref{topol20}) gives the smooth mapping of the space-time hypersurface $S^3$
prescribed by the Euler angles (see Eqs. (\ref{inst108}) with a fixed parameter $r$) into the $\mbox{SU(2)}$ group
space, and the degree of the mapping is equal to $(-1)$.

Write out also the Riemannian curvature 2-form:
\begin{gather}
R^{\alpha}_{(-)}=0,
\nonumber \\
R^1_{(+)}=-\frac{8a^4}{r^6}\left(e^1\wedge e^4+e^2\wedge e^3\right),
\nonumber \\
R^2_{(+)}=\frac{8a^4}{r^6}\left(e^1\wedge e^3-e^2\wedge e^4\right),
\nonumber \\
R^3_{(+)}=\frac{16a^4}{r^6}\left(e^1\wedge e^2+e^3\wedge e^4\right).
\label{inst150}
\end{gather}

\section{The lattice gravity model}

The next step is to adumbrate the model of lattice gravity which is used here.
A detailed description of the model is given in \cite{8}-\cite{10}.

The orientable 4-dimensional simplicial complex and its vertices are designated as $\mK$ and
$a_{\cV}$, the indices ${\cV}=1,2,\dots,\,{\mN}\rightarrow\infty$ and ${\cW}$ enumerate the vertices
and 4-simplices, correspondingly. It is necessary to use
the local enumeration of the vertices $a_{\cV}$ attached to a given
4-simplex: the all five vertices of a 4-simplex with index ${\cW}$
are enumerated as $a_{{\cW}i}$, $i=1,2,3,4,5$. The later notations with extra index  ${\cW}$
indicate that the corresponding quantities belong to the
4-simplex with index ${\cW}$. The Levi-Civita symbol with in pairs
different indexes $\varepsilon_{{\cW}ijklm}=\pm 1$ depending on
whether the order of vertices
$s^4_{\cW}=a_{{\cW}i}a_{{\cW}j}a_{{\cW}k}a_{{\cW}l}a_{{\cW}m}$ defines the
positive or negative orientation of 4-simplex $s^4_{\cW}$.
An element of the group $\Spin(4)$ and an element of the Clifford algebra
\begin{gather}
\Omega_{{\cW}ij}=\Omega^{-1}_{{\cW}ji}=\exp\left(\omega_{{\cW}ij}\right), \quad
\omega_{{\cW}ij}\equiv\frac{1}{2}\sigma^{ab}\omega^{ab}_{{\cW}ij},
\nonumber \\
\hat{e}_{{\cW}ij}=\hat{e}_{{\cW}ij}^{\dag}\equiv e^a_{{\cW}ij}\gamma^a\equiv-\Omega_{{\cW}ij}\hat{e}_{{\cW}ji}\Omega_{{\cW}ij}^{-1}.
\label{discr20}
\end{gather} 
are assigned for each oriented 1-simplex $a_{{\cW}i}a_{{\cW}j}$. The lattice analog of the
action (\ref{inst20}) has the form
\begin{gather}
\mA=\frac{1}{5\times
24}\sum_{\cW}\sum_{i,j,k,l,m}\varepsilon_{{\cW}ijklm}\tr\gamma^5 \times
\nonumber \\
\times\left\{-\frac{2}{l^2_P}\Omega_{{\cW}mi}\Omega_{{\cW}ij}\Omega_{{\cW}jm}
\hat{e}_{{\cW}mk}\hat{e}_{{\cW}ml}\right\}.
\label{discr40}
\end{gather}
This action is invariant relative to the gauge transformations
\begin{gather}
\tilde{\Omega}_{{\cW}ij}=S_{{\cW}i}\Omega_{{\cW}ij}S^{-1}_{{\cW}j},
\quad
\tilde{e}_{{\cW}ij}=S_{{\cW}i}\,e_{{\cW}ij}\,S^{-1}_{{\cW}i}, 
\nonumber \\
 S_{{\cW}i}\in\Spin(4).
\label{discr45}
\end{gather}

It is natural to interpret the quantity
\begin{gather}
l^2_{{\cW}ij}\equiv\frac{1}{4}\,\tr\,(\hat{e}_{{\cW}ij})^2=\sum_{a=1}^4(e^a_{{\cW}ij})^2\sim l_P^2
\label{discr110}
\end{gather}
as the square of the length of the edge $a_{{\cW}i}a_{{\cW}j}$, and the parameter $l_P$ is of the order of the lattice spacing. Thus, the geometric properties
of a simplicial complex prove to be completely defined.

Now, let us show in the limit of slowly varying fields, that the action
(\ref{discr40}) reduces to the continuous gravity action (\ref{inst20}).

Consider a certain  $4D$ sub-complex of complex $\mK$ with the trivial topology of four-dimensional disk. 
Realize geometrically this sub-complex in $\R^4$.  Suppose that the
geometric realization is an almost smooth four-dimensional
surface {\footnote{Here, by an almost smooth surface, we
mean a piecewise smooth surface consisting of flat
four-dimensional simplices, such that the angles between
adjacent 4-simplices tend to zero and the sizes of these
simplices are commensurable.}}. Thus each vertex of the sub-complex  acquires
the coordinates $x^{\mu}$  which are the coordinates of the vertex image in $\R^4$: 
\begin{gather}
x^{\mu}_{{\cW}i}=x^{\mu}_{\cV}\equiv x^{\mu}(a_{{\cW}i})\equiv x^{\mu}(a_{\cV}),
 \qquad \ \mu=1,\,2,\,3,\,4
\label{discr120}
\end{gather}
We stress that these coordinates
are defined only by their vertices rather than by the higher dimension simplices
to which these vertices belong; moreover, the correspondence between the vertices
from the considered subset and the coordinates (\ref{discr120}) is one-to-one.

We have
\begin{gather}
|\,x^{\mu}_{{\cW}i}-x^{\mu}_{{\cW}j}\,|\sim l_P.
\label{discr130}
\end{gather}

It is evident that the four vectors
\begin{gather}
\d x^{\mu}_{{\cW}ji}\equiv x^{\mu}_{{\cW}i}-x^{\mu}_{{\cW}j},
 \quad i=1,\,2,\,3,\,4
\label{discr140}
\end{gather}
are linearly independent and
\begin{gather}
\left\vert
\begin{array}{llll}
\d x^1_{{\cW}m1} \ & \ \d x^2_{{\cW}m1} \ & \ldots & \ \d x^4_{{\cW}m1}\\
\ldots  & \ldots  &  \ldots  & \ldots \\
\d x^1_{{\cW}m4} \ & \ \d x^2_{{\cW}m4} \ & \ldots & \ \d x^4_{{\cW}m4}
\end{array}\right\vert\gtrless0,
\label{discr150}
\end{gather}
depending on whethe the frame $\big(X^{\cW}_{m\,1},\,
\ldots\,,\,X^{\cW}_{m\,4}\,\big)$ is positively or negatively oriented. Here, the differentials of coordinates
(\ref{discr140}) correspond to one-dimensional simplices $a_{{\cW}j}a_{{\cW}i}$, so that,
if the vertex $a_{{\cW}j}$ has coordinates $x^{\mu}_{{\cW}j}$, then the vertex
$a_{{\cW}i}$ has the coordinates $x^{\mu}_{{\cW}j}+\d x^{\mu}_{{\cW}ji}$.

In the continuous limit, the holonomy group elements (\ref{discr20}) are
close to the identity element, so that the quantities $\omega^{ab}_{ij}$
tend to zero being of the order of $O(\d x^{\mu})$.
Thus one can consider the following system of equation for $\omega_{{\cW}m\mu}$
\begin{gather}
\omega_{{\cW}m\mu}\,\d x^{\mu}_{{\cW}mi}=\omega_{{\cW}mi},  \quad
i=1,\,2,\,3,\,4\,.
\label{discr160}
\end{gather}
In this system of linear equation, the indices ${\cW}$ and $m$ are
fixed, the summation is carried out over the index $\mu$, and
index runs over all its values. Since the determinant
(\ref{discr150}) is nonzero, the quantities $\omega_{{\cW}m\mu}$
are defined uniquely. Suppose that a one-dimensional simplex
$X^{\cW}_{m\,i}$ belongs to four-dimensional simplices with indices
${\cW}_1,\,{\cW}_2,\,\ldots\,,\,{\cW}_r$. Introduce the quantity
\begin{gather}
\omega_{\mu}\left(\frac{1}{2}\,(x_{{\cW}m}+
x_{{\cW}i})\,\right)\equiv\frac{1}{r}\,
\bigg\{\omega_{{\cW}_1m\mu}+\,\ldots\,+\omega_{{\cW}_rm\mu}\,\bigg\}\,,
\label{discr170}
\end{gather}
which is assumed to be related to the midpoint of the segment
$[x^{\mu}_{{\cW}m},\,x^{\mu}_{{\cW}i}\,]$. Recall that the coordinates
$x^{\mu}_{{\cW}i}$ as well as the differentials (\ref{discr140})
depend only on vertices but not on the higher dimensional
simplices to which these vertices belong. According to the
definition, we have the following chain of equalities
\begin{gather}
\omega_{{\cW}_1\,mi}=\omega_{{\cW}_2\,mi}= \,\ldots\,=\omega_{{\cW}_r\,mi}\,.
\label{discr180}
\end{gather}
It follows from (\ref{discr140}) and
(\ref{discr160})--(\ref{discr180}) that
\begin{gather}
\omega_{\mu}\left(x_{{\cW}m}+ \frac{1}{2}\,\d x_{{\cW}mi}\,\right)\,\d
x^{\mu}_{{\cW}mi}=\omega_{{\cW}mi}  \,.
\label{discr190}
\end{gather}
The value of the field element $\omega_{\mu}$ in (\ref{discr190}) is uniquely defined by the corresponding
one-dimensional simplex.

Next, we assume that the fields $\omega_{\mu}$ smoothly depend on
the points belonging to the geometric realization of each
four-dimensional simplex. In this case, the following formula is
valid up to $O\big((\d x)^2\big)$ inclusive
\begin{gather}
\Omega_{{\cW}mi}\,\Omega_{{\cW}ij}\,\Omega_{{\cW}jm}=
\exp\left[\frac{1}{2}\,\mR_{\mu\nu}(x_{{\cW}m})\d x^{\mu}_{{\cW}mi}\, \d
x^{\nu}_{{\cW}mj}\,\right]\,,
\label{discr200}
\end{gather}
where
\begin{gather}
\mR_{\mu\nu}=\partial_{\mu}\omega_{\nu}-\partial_{\nu}\omega_{\mu}+
[\omega_{\mu},\,\omega_{\nu}\,]\,.
\label{discr210}
\end{gather}
When deriving formula (\ref{discr200}), we used the
Hausdorff formula.

In exact analogy with (\ref{discr160}), let us write out the following relations
for a tetrad field without explanations
\begin{gather}
\hat{e}_{{\cW}m\mu}\,\d x^{\mu}_{{\cW}mi}=\hat{e}_{{\cW}mi}\,.
\label{discr220}
\end{gather}

Applying formulas (\ref{discr200})--(\ref{discr220}) to the discrete action (\ref{discr40}) and
changing the summation to integration we find that
in the long-wavelength limit the lattice action (\ref{discr40}) transforms into 
continuous action (\ref{inst20}) and any information about lattice is forgotten in the main approximation.

\section{The self-dual solutions to lattice gravity}

Now let us consider the self-dual solution to lattice gravity.
We have the lattice analogue of Eqs. (\ref{inst60}) and (\ref{inst90}): 
\begin{gather}
\delta\mA/\delta\omega^{\alpha}_{(\pm){\cW}mi}=0, 
\label{lat10}
\end{gather}
\begin{gather}
\delta\mA/\delta e^a_{{\cW}mi}=0.
\label{lat20}
\end{gather}
Because of the action (\ref{discr40}) is a homogeneous quadratic function of the
variables $\{e\}$, so   
\begin{gather}
2\mA=\sum_{\{e\}}e^a_{{\cW}mi}
\left(\delta\mA/\delta e^a_{{\cW}mi}\right)=0
\label{lat30}
\end{gather}
on the mass shell according to Euler theorem. Let's impose the additional conditions (compare with Eqs. (\ref{inst95}))
\begin{gather}
\omega^{\alpha}_{(-){\cW}mi}=0.
\label{lat40}
\end{gather}
Combining Eqs. (\ref{lat10}) with index $(-)$ and (\ref{lat40}) we obtain:
\begin{gather}
\frac{\delta\mA}{\delta\omega^{\alpha}_{(-){\cW}mi}}\bigg|_{\omega^{\alpha}_{(-)}=0}=-\frac{2}{l^2_P}
\sum_{{\cW}'}\sum_{j,k,l}\varepsilon_{{\cW}'mijkl}\bigg\{E^{\alpha}_{(-){\cW}'[jkl]}+
\nonumber \\
+\frac16\Big[\big(E^{\alpha}_{(-){\cW}'m[kl]}+E^{\alpha}_{(-){\cW}'m[lj]}+E^{\alpha}_{(-){\cW}'m[jk]}\big)+
\nonumber \\
+\big(m\leftrightarrow i\big)\Big]\bigg\}=0,
\nonumber \\
E^{\alpha}_{*[mkl]}\equiv\frac{1}{3}\bigg(E^{\alpha}_{*m[kl]}+
E^{\alpha}_{*k[lm]}+E^{\alpha}_{*l[mk]}\bigg),
\nonumber \\
E^{\alpha}_{(\pm)*m[kl]}\equiv \mp\left(e^{\alpha}_{*mk}e^4_{*ml}-
e^4_{*mk}e^{\alpha}_{*ml}\right)+
\nonumber \\
+\varepsilon_{\alpha\beta\gamma}e^{\beta}_{*mk}e^{\gamma}_{*ml}.
\label{lat50}
\end{gather}
The index ${\cW}'$ in (\ref{lat50}) enumerates all 4-simplices which contain a marked
1-simplex $a_{{\cW}m}a_{{\cW}i}$. As in continuous case, the system of equations
(\ref{lat10}), (\ref{lat20}) and (\ref{lat40}) is equivalent to the system of equations
(\ref{lat10}) with index $(+)$, (\ref{lat20}), (\ref{lat40}) and (\ref{lat50}).
It will be shown that the square brackets give null equation under the sum (\ref{lat50}).
Therefore only the first term in the parentheses in Eq. (\ref{lat50}) is significant.

Equations (\ref{lat50}) do not fix the quantity $E^{\alpha}_{(-){\cW}'[jkl]}$ completely 
but up to summands of the kind 
\begin{gather}
E^{\alpha}_{(-){\cW}'[jkl]}\longrightarrow 
\nonumber \\
\longrightarrow E^{\alpha}_{(-){\cW}'[jkl]}+
\left(\Psi^{\alpha}_{(-){\cW}jk}+
\Psi^{\alpha}_{(-){\cW}kl}+\Psi^{\alpha}_{(-){\cW}lj}\right),
\label{lat60}
\end{gather}
and the lattice 1-form $\Psi^{\alpha}_{(-){\cW}mk}=-\Psi^{\alpha}_{(-){\cW}km}$ can be varied according to
\begin{gather}
\Psi^{\alpha}_{{(-)\cW}mk}\longrightarrow \Psi^{\alpha}_{(-){\cW}mk}+\left(\phi^{\alpha}_{(-){\cW}k}-\phi^{\alpha}_{(-){\cW}m}\right).
\label{lat70}
\end{gather}
It follows from here that Eqs. (\ref{lat50}) fix no more than  3 additional real-valued parameters
at each vertex of complex, leading to the fixation of gauge subgroup $\Spin(4)_{(-)}$.

Let's prove the given statements. For this end one must check that the equation
\begin{gather}
\sum_{{\cW}'}\bigg\{\sum_{j,k,l}\varepsilon_{{\cW}'mijkl}\big(\Psi^{\alpha}_{(-){\cW}'jk}+
\Psi^{\alpha}_{(-){\cW}'kl}+
\nonumber \\
\Psi^{\alpha}_{(-){\cW}'lj}\big)\bigg\}=0
\label{lat80}
\end{gather}
is satisfied identically. It is evident that the braces in Eq. (\ref{lat80}) vanishes identically at each fixed value of index
${\cW}'$ if $\Psi^{\alpha}_{(-){\cW}'mk}=\left(\phi^{\alpha}_{(-){\cW}'k}-\phi^{\alpha}_{(-){\cW}'m}\right)$.

Consider two adjacent positively oriented 4-simplices
\begin{gather}
s^4_{\cW}=a_{{\cW}m}a_{{\cW}i}a_{{\cW}j}a_{{\cW}k}a_{{\cW}l}, 
\nonumber \\
s^4_{{\cW}'}=a_{{\cW}'m}a_{{\cW}'i}a_{{\cW}'k}a_{{\cW}'j}a_{{\cW}'l'},
\nonumber \\
a_{{\cW}m}=a_{{\cW}'m}, \quad a_{{\cW}i}=a_{{\cW}'i}, \quad a_{{\cW}j}=a_{{\cW}'j},
\nonumber \\ 
a_{{\cW}k}=a_{{\cW}'k}, \quad a_{{\cW}l}\neq a_{{\cW}'l'},
\label{lat90}
\end{gather}
so that
\begin{gather}
\varepsilon_{{\cW}mijkl}=\varepsilon_{{\cW}'mikjl'}=1.
\label{lat100}
\end{gather}
Select from the sum (\ref{lat80}) two summands corresponding to the 4-simplices (\ref{lat90}):
\begin{gather}
\sum_{j,k,l}\varepsilon_{{\cW}mijkl}\big(\Psi^{\alpha}_{(-){\cW}jk}+
\Psi^{\alpha}_{(-){\cW}kl}+\Psi^{\alpha}_{(-){\cW}lj}\big)+
\nonumber \\
+\sum_{j,k,l'}\varepsilon_{{\cW}'mikjl'}\big(\Psi^{\alpha}_{(-){\cW}'kj}+
\Psi^{\alpha}_{(-){\cW}'jl'}+\Psi^{\alpha}_{(-){\cW}'l'k}\big).
\label{lat110}
\end{gather}
We see that 
the quantities $\Psi^{\alpha}_{(-){\cW}mk}$ belonging to the common 3-simplices of the adjacent 4-simplices cancel on
in (\ref{lat110}) as a consequence of Eqs. (\ref{lat100}) and $\left(\Psi^{\alpha}_{(-){\cW}jk}+\Psi^{\alpha}_{(-){\cW}'kj}\right)=0$. Note that each quantity $\Psi^{\alpha}_{(-){\cW}'jk}$ in parentheses in (\ref{lat80})
belongs to the 3-simplex which is common to {\it two} adjacent 4-simplices. 
If not, the cavities and  boundaries would be in the simplicial complex, but such complexes are not
considered here. Thus, the sum (\ref{lat80}) is equal to zero identically \footnote{Hence the
statement that the square brackets give null equation under the sum (\ref{lat50}) is proved.}. 
This means that the system of equations
(\ref{lat10}) with index $(+)$, (\ref{lat20}), (\ref{lat40}) and (\ref{lat50}) is self-consistent
and it fixes in part the gauge as well as in continuous case. According to the Eqs. (\ref{lat20}) and
(\ref{lat30}) the action of any solution of this system of equations is equal to zero.

\section{Lattice analogue of the Eguchi-Hanson solution}

Now proceed to study the lattice analog of the Eguchi-Hanson solution. But it is impossible to give the irregular lattice solution in an explicit form in contrast to
the continuous case.
Thus, the problem reduces to the solution existence proof and finding of its asymptotics.

Suppose that the complex $\mK$ is finite but it contains gillion  simplices, so that the number of
vertices $\infty>\mN\ggg1$. Suppose also that $\mK$ can be considered as a triangulation of a part of $\R^4$
and $\partial\mK\approx S^3$. It is assumed that in a wide vicinity of $\partial\mK$
the long-wavelength limit is valid and the continuous solution (\ref{inst110}), (\ref{inst140}) approximates
correctly the exact lattice solution and the  hypersurface $\partial\mK$ is given by
the Eq. $r=R=\Const\longrightarrow\infty$. 
Denote by $\mk\subset\mK$ a finite sub-complex  containing the centre of instanton with the
boundary $\partial\mk\approx S^3$.
Evidently, the Euler characteristics $\chi({\mk})=\chi({\mK})=1$.
We have the exact lattice equivalents of the instanton boundary conditions (\ref{inst143}) and (\ref{inst147}):
\begin{gather}
\frac{1}{12\cdot4!}\sum_{{\cal S}(\partial\mK)}\sum_{ijkm}\varepsilon_{{\cal S}ijkm}\tr
\big(\Omega_{(+){\cal S}jm}^{-1}\Omega_{(+){\cal S}ji}\big)\times
\nonumber \\
\times\big(\Omega_{(+){\cal S}km}^{-1}\Omega_{(+){\cal S}kj}\big)\big(\Omega_{(+){\cal S}im}^{-1}\Omega_{(+){\cal S}ik}\big)=-2\pi^2,
\label{instlat10}
\end{gather}
\begin{gather}
\sum_{{\cal S}(\partial\mk)}\sum_{ijkm}\varepsilon_{{\cal S}ijkm}\tr
\big(\Omega_{(+){\cal S}jm}^{-1}\Omega_{(+){\cal S}ji}\big)\times
\nonumber \\
\times\big(\Omega_{(+){\cal S}km}^{-1}\Omega_{(+){\cal S}kj}\big)\big(\Omega_{(+){\cal S}im}^{-1}\Omega_{(+){\cal S}ik}\big)=0.
\label{instlat20}
\end{gather}
Here the indices ${\cal S}(\partial\mK)$ and ${\cal S}(\partial\mk)$ enumerate 3-simplices on the boundaries 
$\partial\mK$ and $\partial\mK$, correspondingly,
and the Levi-Civita symbol $\varepsilon_{{\cal S}ijkm}=\pm 1$ depending on
whether the order of vertices
$s^3_{{\cal S}}=a_{{\cal S}i}a_{{\cal S}j}a_{{\cal S}k}a_{{\cal S}m}$ defines the
positive or negative orientation of this 3-simplex.

Since the long-wavelength limit is valid as $r\longrightarrow\infty$, one can use
the instanton solution for the dynamic variables (\ref{inst110})-(\ref{inst140}) in this region:
\begin{gather}
\Omega_{(+){\cW}mi}\approx1+\frac{i\sigma^{\alpha}}{2}\omega^{\alpha}_{(+)\mu}\d x^{\mu}_{{\cW}mi}.
\nonumber
\end{gather}
Therefore the sum in (\ref{instlat10}) transforms into integral (\ref{inst143})
 with the only difference that now the angle $\psi$ varies in the interval (\ref{inst130}),
 and so the boundary condition (\ref{instlat10}) is just.

To implement the boundary condition (\ref{instlat20}) we suggest the following solutions on the sub-complex $\mk$.

Consider the following solutions of Eqs. (\ref{lat10}), (\ref{lat20}) and (\ref{lat40})  on $\mk$:
\begin{gather}
\Omega_{(+){\cW}ij}=-1,   \quad  \Omega_{(-){\cW}ij}=1, \quad s^4_{\cW}\in\mk.
\label{instlat30}
\end{gather}
From here it follows that
\begin{gather}
\Omega_{(+){\cW}mi}\Omega_{(+){\cW}ij}\Omega_{(+){\cW}jm}=-1 \quad \mbox{on} \quad \mk.
\label{instlat40}
\end{gather}
The equalities (\ref{instlat40}) hold true if we perform a gauge transformation (see (\ref{discr45}))
with $S_{(+){\cW}i}=\pm1$, $S_{(-){\cW}i}=1$. 
The boundary condition (\ref{instlat20}) is true for each configuration obtained in such a way.

Obviously, 1-form $\hat{e}_{{\cW}ij}$ can be considered as a 1-cochain on complex, and the
quantity
\begin{gather}
E^{ab}_{{\cW}m[kl]}=(E^{\alpha}_{(+){\cW}m[kl]}, \ (E^{\alpha}_{(-){\cW}m[kl]})=
\nonumber \\
=\varepsilon_{abcd}e^c_{{\cW}mk}e^d_{{\cW}ml},
\label{instlat50}
\end{gather}
is a  2-cochain which is the superposition of the exterior products of 1-cochains $e^a_{{\cW}ij}$.

It may be verified (compare with (\ref{lat50})) that the left hand side of Eq. (\ref{lat10}) is nothing but the exterior lattice derivative of a 2-cochain (\ref{instlat50}),
and Eq. (\ref{lat10}) implies that the derivative is equal to zero in the case of (\ref{instlat40}),
i.e. the quantity (\ref{instlat50}) is a cocycle \footnote{It was proved to be the case for the
quantity $E^{\alpha}_{(-){\cW}[mkl]}$ (see (\ref{lat50})). The corresponding provement
for the quantity $E^{\alpha}_{(+){\cW}[mkl]}$ is the same.}: 
\begin{gather}
\sum_{{\cW}'}\sum_{j,k,l}\varepsilon_{{\cW}'mijkl}E^{ab}_{{\cW}'[jkl]}=0,
\nonumber \\
a_{{\cW}m}a_{{\cW}i}\in s^4_{\cW'}, \quad  a_{{\cW}m}a_{{\cW}i}\in\mk, \quad  a_{{\cW}m}a_{{\cW}i}\notin\partial\mk.
\label{instlat60}
\end{gather}
Evidently, Eqs. (\ref{instlat60}) are satisfied for
\begin{gather}
e^a_{{\cV}_1{\cV}_2}=\phi^a_{{\cV}_2}-\phi^a_{{\cV}_1}, \quad a_{{\cV}_1}a_{{\cV}_2}\in\mk, \quad  a_{{\cV}_1}a_{{\cV}_2}\notin\partial\mk.
\label{instlat70}
\end{gather}
where $\phi^a_{{\cV}}$ is any scalar field on ${\mk}$. This statement is checked easily by a direct
calculation and it follows from the fact that  the cochain (\ref{instlat50}) is the
superposition of the exterior products of exact 1-forms in the case (\ref{instlat70}).

Since the second cohomology group $H^2(\mk)=0$, so the cocycle (\ref{instlat50}) is a lattice coboundary:
\begin{gather}
E^{ab}_{{\cW}m[kl]}=\left(\Psi^{ab}_{{\cW}mk}-\Psi^{ab}_{{\cW}ml}\right),
\nonumber \\
\Psi^{ab}_{{\cW}mk}=-\Psi^{ab}_{{\cW}km}.
\label{instlat80}
\end{gather}
Eqs. (\ref{instlat60}) fix no more than 6 real numbers at each vertex of the sub-complex $\mk$ for the reason that the lattice 1-form $\Psi^{ab}_{{\cW}mk}$ is determined up to the lattice gradient $(\phi^{ab}_{{\cW}k}-\phi^{ab}_{{\cW}m})$.
So it follows from Eqs. (\ref{instlat30}) (just like as in the case of (\ref{lat40}))  that now not only the gauge sub-group $\Spin(4)_{(-)}$ is broken but also the gauge group $\Spin(4)$ is  broken to 
on $\mk\setminus\partial\mk$ almost wholly except for the center of the sub-group $\SU(2)_{(+)}$  (see Eqs.  (\ref{lat60})-(\ref{lat110})).

From here and throughout the following discussions the pairs of indices $({\cV}_1{\cV}_2)$ enumerate 1-simplexes $a_{{\cV}_1}a_{{\cV}_2}\in\mK$.

The action (\ref{discr40}) is equal to zero identically for the configurations (\ref{lat40}), (\ref{instlat40}) 
and any values of $e^a_{{\cV}_1{\cV}_2}$, where 1-simplex $a_{{\cV}_1}a_{{\cV}_2}\in\mk$,
$a_{{\cV}_1}a_{{\cV}_2}\notin\partial\mk$. Therefore, Eqs. (\ref{lat20}) are satisfied automatically in this case,
they do not give any constraint onto the corresponding 1-forms $e^a_{{\cV}_1{\cV}_2}$ in addition to Eqs. (\ref{instlat80}):
\begin{gather}
\delta\mA/\delta e^a_{{\cV}_1{\cV}_2}\equiv0, \quad a_{{\cV}_1}a_{{\cV}_2}\in\mk, \quad
a_{{\cV}_1}a_{{\cV}_2}\notin\partial\mk.
\label{instlat90}
\end{gather}

Now let us prove that the lattice configuration of the variables $\left\{\omega^{ab}_{{\cV}_1{\cV}_2(\mbox{inst})}, \ 
e^a_{{\cV}_1{\cV}_2(\mbox{inst})}\right\}$ satisfying the system of equations 
(\ref{lat10}), (\ref{lat20}), (\ref{lat40}) and constraints (\ref{instlat10}), (\ref{instlat20}) does exist.
The configuration is the lattice analogue of the Eguchi-Hanson  continuous instanton.

The constraint (\ref{instlat20}) is realized evidently by taking the variables $\Omega_{(+){\cV}_1{\cV}_2}$
on the sub-complex $\mk$ as in (\ref{instlat30}). We believe also that the relations (\ref{instlat70}) are valid.
The constraint (\ref{instlat10}) also is realized evidently by taking 
\begin{gather}
\Omega_{(+){\cal S}mi}=\exp\left\{\frac{i\sigma^{\alpha}}{2}\omega^{\alpha}_{(+)\mu}\d x^{\mu}_{{\cal S}mi}\right\}\approx
\nonumber \\
\approx1+\frac{i\sigma^{\alpha}}{2}\omega^{\alpha}_{(+)\mu}\d x^{\mu}_{{\cal S}mi}
\label{instlat100}
\end{gather}
on the boundary
$\partial\mK$, where the field $\omega^{\alpha}_{(+)\mu}$ is given by Eq. (\ref{topol10}).

Firstly let's consider  $\mA_g$ as a constrained-free functional (\ref{discr40}) on sub-complex $(\mbox{int}{\mK})\setminus\mk$.
Each of the elements $\Omega_{{\cV}_1{\cV}_2}$ is a smooth matrix function
on the sum of two unit 3D spheres $\left(S^3_{(+)({\cV}_1{\cV}_2)}\cup S^3_{(-)({\cV}_1{\cV}_2)}\right)$. Therefore
the action $\mA_g$ is a smooth real function on the compact manifold without boundary
\begin{gather}
{\cal{C}}=\bigcup_{({\cV}_1{\cV}_2),\,\,a_{{\cV}_1}a_{{\cV}_2}\subset(\mbox{int}{\mK})\setminus\mk}\left(S^3_{(+)({\cV}_1{\cV}_2)}\cup S^3_{(-)({\cV}_1{\cV}_2)}\right)
\label{instlat110}
\end{gather}
for any fixed set of values $\{e\}$. So there exist the points $p_{\xi}\in{\cal{C}}$ at which the action $\mA_g$ is extremal
and therefore
\begin{gather}
\partial\mA_g/\partial\omega^{\alpha}_{(\pm){\cV}_1{\cV}_2}\big|_{p_{\xi}\in{\cal{C}}}=0.
\label{instlat120}
\end{gather}
Further Eqs.(\ref{instlat120}) are assumed to be true.

Recall that the sets of equations (\ref{lat10}) with the sign $(-)$ and (\ref{lat40}) are equivalent to the set of
equations  (see Eqs. (\ref{lat50}))
\begin{gather}
\Phi^{\alpha}_{({\cV}_1{\cV}_2)}\equiv -\frac{l^2_P}{2}
\frac{\delta\mA}{\delta\omega^{\alpha}_{(-){\cV}_1{\cV}_2}}\bigg|_{\omega_{(-)}=0}=
\nonumber \\
=\sum_{{\cW}'}\sum_{j,k,l}\varepsilon_{{\cW}'{\cV}_1{\cV}_2jkl}E^{\alpha}_{(-){\cW}'[jkl]}=0
\label{instlat130}
\end{gather}
and Eqs. (\ref{lat40}).
Here the quantities $a_{{\cW}m}a_{{\cW}i}$ and
$\varepsilon_{{\cW}'mijkl}$ are  renamed as $a_{{\cV}_1}a_{{\cV}_2}$  and $\varepsilon_{{\cW}'{\cV}_1{\cV}_2jkl}$,
correspondingly. Further the set of equations (\ref{instlat130}) on ${\mK}\setminus\mbox{int}\,{\mk}$ is considered as the set of constraints.

Let's resolve the constraints (\ref{instlat130})  on the sub-complex $\left({\mK}\setminus\mbox{int}\,{\mk}\right)$ relative to a subset of variables 
$\{\mg_q\},\,q=1,\ldots,Q\sim3{\mN}$,
describing the gauge transformations $\Spin(4)_{(-)}$. 
We denote by $\{e'\}$ the rest of variables $\{e\}$. So we have on the sub-complex $\left({\mK}\setminus\mbox{int}{\mk}\right)$:

(i) the set of variables $\{e\}$ is equivalent to the set of
variables $\{e',\,\mg\}$,

(ii) the hypersurface (\ref{instlat130}) is represented as the set of 
\linebreak
analytical functions $\mg(e')$ and it is designated
as ${\cal E}$.

The hypersurface ${\cal E}$  exhibits the following properties:

a) The manifold ${\cal E}$ is boundaryless. 

This property follows from the  implicit function theorem.

b) The hypersurface ${\cal E}$ is a  closed set.

This property follows from the fact that any accumulation point of ${\cal E}$ in the space $\{e\}$
belong to the hypersurface due to analyticity of the constraints (\ref{instlat130}).

Denote by $S^H$ the hypersphere
\begin{gather}
\sum_{({\cV}_1{\cV}_2),\,\,a_{{\cV}_1}a_{{\cV}_2}\subset{\mK}\setminus(\mbox{int}\mk)}\left(e^a_{{\cV}_1{\cV}_2}\right)^2=
l_P^2{\mN},
\label{instlat140}
\end{gather}
and
\begin{gather}
{\mC}=S^H\cap{\cal E}, \quad \partial{\mC}=0.
\label{instlat150}
\end{gather}
The manifold ${\mC}$ is compact and boundaryless.

It follows from here that the points $q_{\zeta}\in{\mC}$  exist
at which the action  ${\mA}_g(e',\,\mg)$  is extremal 
and so
\begin{gather}
\partial{\mA}_g(e',\,\mg)/\partial e'\big|_{q_{\zeta}\in{\mC}}=0.
\label{instlat160}
\end{gather}
We have also
\begin{gather}
\partial{\mA}_g(e',\,\mg)/\partial\mg=0.
\label{instlat170}
\end{gather}
The set of Eqs. (\ref{instlat170}) follow from the fact that variations of variables $\{\mg\}$ lead to variations of
the constraints (\ref{instlat130}) only. But  herein the corresponding variations of the action $\mA$ are equal to zero
due to Eqs. (\ref{lat40}) and since the contribution of the constraints (\ref{instlat130}) into the action looks like as
\begin{gather}
\Delta\mA\sim\sum\omega_{(-)}\Phi_{(-)}.
\nonumber
\end{gather}

The totality of equations (\ref{instlat160}) and (\ref{instlat170}) is equivalent to the system of equations
\begin{gather}
\partial{\mA}_g(e)/\partial e\big|_{S^H}=0.
\label{instlat180}
\end{gather}

Resolve the constraint (\ref{instlat140}) relative to
$e^1_{{\cV}_{{\mN}-1}{\cV}_{{\mN}}}$:
\begin{gather}
e^1_{{\cV}_{{\mN}-1}{\cV}_{{\mN}}}=\pm f(\ldots), \quad a_{{\cV}_{{\mN}-1}}a_{{\cV}_{{\mN}}}\in\partial{\mK},
\nonumber \\
f=\sqrt{l_P^2{\mN}-\sum{}^{\prime}\left(e^a_{{\cV}_1{\cV}_2}\right)^2}.
\label{instlat190}
\end{gather}
Here the symbol $\sum{}^{\prime}$ is the same as in (\ref{instlat140}) only that
the variable $e^1_{{\cV}_{{\mN}-1}{\cV}_{{\mN}}}$ is excluded. It follows from (\ref{instlat180}), (\ref{instlat190}):
\begin{gather}
\frac{\partial{\mA}_{g}}{\partial e^a_{{\cV}_1{\cV}_2}}\pm
\bigg(\frac{\partial{\mA}_{g}}{\partial e^1_{{\cV}_{{\mN}-1}{\cV}_{{\mN}}}}\cdot
\frac{\partial f}{\partial e^a_{{\cV}_1{\cV}_2}}\bigg)=0.
\label{instlat200}
\end{gather}

There is the estimation
\begin{gather}
l_P^2{\mN}\sim R^4.
\label{instlat210}
\end{gather}

According to (\ref{discr40}) and the Eguchi-Hanson solution (\ref{inst150})
\begin{gather}
\frac{\partial{\mA}_g}{\partial e^1_{{\cV}_{{\mN}-1}{\cV}_{{\mN}}}}\sim{\mR}^{ab}\big|_{\mbox{ not far from}\,\,\partial{\mK}}
\sim R^{-6}\sim{\mN}^{-3/2},
\label{instlat220}
\end{gather}
and therefore 
\begin{gather}
\frac{\partial{\mA}_{g}}{\partial e^a_{{\cV}_1{\cV}_2}}\sim R^{-6}.
\label{instlat230}
\end{gather}
Since the number of 1-simplexes $a_{{\cV}_1{\cV}2}\subset\mK$ is of the order of $R^4$, so the estimation
\begin{gather}
 \sum_{({\cV}_1{\cV}_2)}e^a_{{\cV}_1{\cV}_2}\bigg(\frac{\partial{\mA}_g}{\partial e^1_{{\cV}_{{\mN}-1}{\cV}_{{\mN}}}}\cdot
\frac{\partial f_{{\mN}}}{\partial e^a_{{\cV}_1{\cV}_2}}\bigg)\sim R^{-2},
\label{instlat240}
\end{gather}
holds. Now due to (\ref{instlat240}) and (\ref{lat30}) we have 
\begin{gather}
2\mA_g=\sum_{({\cV}_1{\cV}_2)}
 e^a_{{\cV}_1{\cV}_2}\frac{\partial{\mA}_{g}}{\partial e^a_{{\cV}_1{\cV}_2}}=
\nonumber \\
=\mp\sum_{({\cV}_1{\cV}_2)}e^a_{{\cV}_1{\cV}_2}\bigg(\frac{\partial{\mA}_g}{\partial e^1_{{\cV}_{{\mN}-1}{\cV}_{{\mN}}}}\cdot
\frac{\partial f_{{\mN}}}{\partial e^a_{{\cV}_1{\cV}_2}}\bigg)\sim  R^{-2}.
\label{instlat250}
\end{gather}

With the help of Eqs. (\ref{instlat230}) and (\ref{instlat250}) we obtaine the final result:
\begin{gather}
\frac{\partial{\mA}}{\partial e^a_{{\cV}_1{\cV}_2}}\longrightarrow 0, \quad \mA\longrightarrow 0
\nonumber \\
 \mbox{as} \quad {\mN}\longrightarrow\infty, \quad R\longrightarrow\infty.
\label{instlat260}
\end{gather}

So, it is proved the existence of  simultaneous solution of equations of Eqs. (\ref{lat10}), (\ref{lat20}), (\ref{lat40}) with boundary conditions (\ref{instlat30}), (\ref{instlat100}) on infinite complex $\mK\approx \R^4$.

\section{Conclusion}

Aforesaid means
that the lattice analogue of the Eguchi-Hanson self-dual solution to continuous Euclidean Gravity does
exist for the complex $\mK\approx \R^4$.

It is crucially important that the problem of possible singularities for the curvature tensor
does not exist on the lattice gravity. In continuous gravity for the $\psi$ range (\ref{inst130}),
the manifold would have "cone-tip" singularities at $r=a$; this implies the necessity of
delta-functions in the curvature at $r=a$ (see \cite{2,3}). But 
delta-functions transforms into Kronecker symbol which is of the order of unity in discrete mathematics.
The same is true with respect to the lattice analogue of the curvature tensor
$\Omega_{{\cW}mi}\Omega_{{\cW}ij}\Omega_{{\cW}jm}\sim\pm1$. This is the reason why one can take the
range of angles (\ref{inst130}) in a lattice gravity.  Moreover, all lattice equations are satisfied
and the action for the instanton solution is equal to zero.

Thus, the setting of the problem for finding a self-dual solution to lattice Euclidean gravity
is as follows: one must solve (in anti-instanton case) the difference lattice system of equations 
(\ref{lat10}) with indices $(+)$, (\ref{lat20}), (\ref{lat40}), the constraints  (\ref{instlat130}), and the
boundary conditions (\ref{instlat30}), (\ref{instlat100}).

Here some questions are not enough clear or remain unclear.

1) Is the offered solution with $\chi({\mK})=1$ stable or it can be contracted smoothly into the trivial one?

The answer to this question seems to be as follows.

Let's consider a smooth path in the configuration space
\begin{gather}
\Omega_{{\cV}_1{\cV}_2}(t), 
\quad e^{a}_{{\cV}_1{\cV}_2}(t), \quad
0\leq t\leq 1, 
\nonumber
\end{gather}
such that 
\begin{gather}
\Omega_{{\cV}_1{\cV}_2}(0)=\Omega_{{\cV}_1{\cV}_2(\mbox{inst})}, \quad
e^{a}_{{\cV}_1{\cV}_2}(0)=e^a_{{\cV}_1{\cV}_2(\mbox{inst})},
\nonumber \\
\Omega_{{\cV}_1{\cV}_2}(1)=1, \quad e^{a}_{{\cV}_1{\cV}_2}(1)=\phi^a_{{\cV}_2}-\phi^a_{{\cV}_1}.
\label{C10}
\end{gather}
Here the lower index $(\mbox{inst})$ designates the discussed self-dual solution.
The field configuration $\left\{\Omega_{{\cV}_1{\cV}_2}(1), \ e^{a}_{{\cV}_1{\cV}_2}(1)\right\}$ is
a trivial solution of Eqs. (\ref{lat10}),  (\ref{lat20}). Evidently, the {\it global} continuous description of this trivial solution is possible.  Nevertheless, the considered instanton contraction scenario seems to be
unsatisfactory since to do this, one must contract topologically non-trivial connection elements (see (\ref{instlat10}) 
and (\ref{instlat100})) into unit in the  infinite space-time.

Another instanton contraction scenario which is described in continual limit by limit process
$a\longrightarrow0$ and in the lattice case by reducing the sub-complex $\mk$ up to its disappearance
also seems to be unsatisfactory. Indeed, in this case we would have resulted in the failure of the 
Chern-Gauss-Bonnet theorem in $\R^4$. So, this scenario is also impossible, it can be considered as
a decreasing of the instanton scale $a$ up to a lattice scale.

Therefore, the considered instanton solution for the case $\chi({\mK})=1$ seems to be stable.

2) The case $\chi({\mk})=(2k+1), \ k=1,2,\ldots$ and $\partial{\mk}=\mS\approx S^3$ is interesting, but it is not considered here. So, the question remains unanswered: for what values of $\chi({\mK})$ the lattice self-dual solution does exist and it would be stable?

\begin{acknowledgments}

I thank A. Maltsev for many consultations. The work has been supported by the RScF grant 16-12-10151.

\end{acknowledgments}

\end{document}